\documentclass[conference]{IEEEtran}
\usepackage{epsfig,cite}
\usepackage{graphicx}
\usepackage{color}
\usepackage{amssymb,amsmath,mathtools, mathrsfs}
\usepackage{enumitem}
\usepackage{graphicx}
\usepackage{epstopdf}

\definecolor{blue}{rgb}{0,0,1}
\definecolor{darkgreen}{rgb}{0,.5,0}
\definecolor{darkred}{rgb}{.5,0,0}

\newtheorem{theorem}{Theorem}
\newtheorem{proposition}{Proposition}

\newtheorem{rem}{Remark}

\def\levy{L\'evy }

\def\ith{$i^{\text{th}}$}

\def\BinomDist{{\mathscr{B}}}
\def\PoissDist{{\mathscr{P}}}
\def\Xprim{X^\prime}
\def\xprim{x^\prime}
\def\Yprim{Y^\prime}
\def\yprim{y^\prime}

\DeclareMathOperator\erfc{erfc}
\DeclareMathOperator\erfcinv{erfcinv}

\IEEEoverridecommandlockouts

\begin{document}
	\bstctlcite{IEEEexample:BSTcontrol}
	
	\title{Capacity of Molecular Channels with Imperfect Particle-Intensity Modulation and Detection}
	
	\author{\IEEEauthorblockN{Nariman Farsad\IEEEauthorrefmark{1}, 
			Christopher~Rose\IEEEauthorrefmark{2}, 
			Muriel~M\'edard\IEEEauthorrefmark{3}, and Andrea~Goldsmith\IEEEauthorrefmark{1}}
		\IEEEauthorblockA{\IEEEauthorrefmark{1}EE, Stanford University,~~~~\IEEEauthorrefmark{2}School of Engineering, Brown University,~~~~\IEEEauthorrefmark{3}EECS, MIT} 
		\vspace{-1.0cm}
	\thanks{This research was supported in part by the NSF Center for Science of Information (CSoI) under grant CCF-0939370.}
	}
	\vspace{-0.8cm}
	
	\maketitle

	\newcommand{\note}[1]{{}}
	\newcommand{\notenf}[1]{{}}
	
	\begin{abstract}
		This work introduces the particle-intensity channel (PIC) as a model for molecular communication systems and characterizes the properties of the optimal input distribution and the capacity limits for this system. In the PIC, the transmitter encodes information, in symbols of a given duration, based on the number of particles released, and the receiver detects and decodes the message based on the number of particles detected during the symbol interval. In this channel, the transmitter may be unable to control precisely the number of particles released, and the receiver may not detect all the particles that arrive. We demonstrate that the optimal input distribution for this channel always has mass points at zero and the maximum number of particles that can be released. We then consider diffusive particle transport, derive the capacity expression when the input distribution is binary, and show conditions under which the binary input is  capacity-achieving. In particular, we demonstrate that when the transmitter cannot generate  particles at a high rate, the optimal input distribution is binary.
	\end{abstract}
	
	\vspace{-0.3cm}
	\section{Introduction}
	\vspace{-0.2cm}
	Molecular communication (MC) conveys information using small particles' time of release, number of release, and/or type \cite{far16ST}. These particles travel from the transmitter to the receiver where they are detected and the message decoded. The transport process is typically random, and introduces uncertainty about the time of particle release and even the number of particles released during a given symbol interval.
	
	One approach to understanding the capacity limits of molecular channels investigated in prior work assumes information is encoded in the time of particle release. Such channels are called molecular timing channels (MTCs). In particular, the additive inverse Gaussian noise channel is presented in \cite{sri12,li14}, and upper and lower bounds on capacity are derived. These work assumed a system where information is encoded in the release time of a single particle, while in \cite{rose16}, molecular timing channels where information is encoded via the release times of multiple particles are considered, and upper and lower bounds on capacity presented. Reference \cite{far16ISIT} presents a MTC, where particles decay after a finite interval, and derives upper and lower bounds on the capacity of this channel.
	
	Another approach to MC encodes information through the number of particles released at the transmitter and decodes based on the number of particles that arrive at the receiver during the symbol interval. We focus on this type of modulation scheme and call it {\em particle-intensity modulation} (PIM)\footnote{This has been called the concentration-shift-keying or the amplitude-modulation in previous work. However, we believe PIM captures the physical properties of this system and its relation to the optical intensity modulation.}. Different channel coding schemes are compared for the MC systems that employ PIM in \cite{Lu15}. In \cite{ein2011, tah15}, this concentration-based channel is considered with a receiver equipped with ligand receptors. The process of molecule reception of a ligand receptor is modeled as a Markov chain and the capacity in bits per channel use is analyzed. The results are extended to multiple access channels in \cite{ami15PtoP}. In \cite{gha15}, a binomial distribution is used to model a system where the transmitter can perfectly control the release of particles and the receiver can perfectly detect the number of particles that arrive.  It is assumed that the channel has finite memory and particle transport is assisted by flow. Using this model, bounds on the capacity are derived, and the capacity for different memory lengths is analyzed. Reference \cite{ami15} assumes that the channel input is the rate of particle release. The channel is represented as a Poisson channel with finite memory, and upper and lower bounds on capacity per channel use are presented. 
	%
	%
	%
	%

	In this paper, molecular channels with imperfect PIM at the transmitter and imperfect detection at the receiver are considered. Specifically, we consider the case where the sender cannot perfectly control the number of particles that are released, and the destination may not detect all the particles that arrive. We assume that the duration of the symbol is long enough such that particles from one symbol do not impinge on future symbols. This model is reasonable if particles diffuse beyond the receptor or disappear in some other fashion, for instance through degradation \cite{guo16}. Under this assumption, the system is memoryless. Finally, we assume that the particles can be generated at a constant fixed rate at the transmitter. For this model, we show that the system can be represented with a channel model similar to the binomial channel \cite{kom01} with two differences. Firstly, the {\em channel input} is discrete instead of continuous. Second, the size of the symbol set, which depends on the maximum number of particles that can be released by the transmitter, changes as a function of symbol duration. This is because of the assumption that the particles are generated at a constant rate at the transmitter. In this work we call this channel the {\em particle-intensity channel (PIC)}. 
	%
	%
	%
	%
	
	%
	%
	%
	%
	
	We define the capacity of this channel in bits per second and as a function of symbol duration. It is shown that the optimal input distribution, for any symbol duration, always has mass points at zero and the maximum number of particles that can be released by the transmitter. This maximum number depends on the rate of particle generation and the symbol duration. One of the most prominent modes of particle transport in the literature is diffusive transport \cite{far16ST}. In this scheme, the released particles would follow a random Brownian path to the receiver. For diffusive particle transport, the capacity of the binary input PIC is derived, where the 0 is represented by sending nothing and 1 by sending the maximum number of particles. We show the conditions under which the binary input is a capacity-achieving distribution. In particular, for the case where the transmitter cannot generate particles at a high rate, an optimal input is binary.

	The rest of this paper is organized as follows. In Section \ref{sec:model} we present the system model and the capacity expression. Then in Section \ref{sec:OptimalInputBCap} we investigate characteristics of the optimal input distribution  and we derive the capacity of the binary input diffusion-based PIC. We present numerical results in Section \ref{sec:results}, and in Section \ref{sec:conclusion} we discuss the results.      
	%
	%
	%
	%
	
	\vspace{-0.2cm}
	\section{System Model and Capacity Formulation}
	\label{sec:model}
	
	\vspace{-0.1cm}
	\subsection{The Particle Intensity Channel} \label{subsec:PIchan}
	
	\vspace{-0.1cm}
	We consider an MC channel in which information is modulated through the PIM, i.e., the number of particles {\em simultaneously released} by the transmitter. The particles themselves are assumed to be {\em identical and indistinguishable} at the receiver, and no other properties (such as the time-of-release) are used for encoding information. The receiver then counts the number of particles that arrive during the symbol duration for decoding the information. The particles that are released by the transmitter travel to the receiver through some random propagation mechanism (e.g., diffusion). We assume that the particles {\em travel independently of each other}, and are {\em detected independently of each other}. This is a reasonable model that has been used in many previous works \cite{far16ST}. 
	%
	%
	%
	%
	%
	%
	%
	%
	%

	The channel is used in a time-slotted fashion, where $\tau$ is the {\em symbol duration}. We define a parameter $\lambda$ as a constant fixed {\em rate} at which the transmitter can {\em generate particles}. Note that in this case the maximum number of particles, $m_\tau$,  that could be released by the transmitter in one channel use can change as a function of symbol duration according to
	\vspace{-0.1cm}
	\begin{align}
	m_\tau = \left\lfloor \lambda\tau \right\rfloor.
	\end{align}
	\vspace{-0.5cm}
	
	We assume that particles are released {\em instantly and simultaneously} at the beginning of the symbol interval. Let $X \mspace{-3mu} \in \mspace{-3mu} \mathcal{X}=\{0,1,\cdots,m_\tau\}$ denote the number of particles the transmitter intends to release, where $\mathcal{X}$ is the input symbol set. Note that $\lambda$ constrains the cardinality of the symbol set $|\mathcal{X}|$. We let $Y$ denote the total number of particles that are detected at the receiver. 
	
	As particles may not dissipate over the symbol duration, particles left over from prior symbol intervals could interfere with detection during the current channel use. Such intersymbol interference (ISI) must be incorporated into deriving the channel capacity, which can be quite challenging, particularly for MC. Therefore, to make the problem more tractable, we assume that particles with transit times exceeding $\tau$ are somehow inactivated.  That is, particles are assumed to have a finite lifetime of duration $\tau$. This feature seems reasonable since particles could be rendered undetectable either naturally or by design (via denaturing or gettering/enzyme reactions \cite{guo16}). With this assumption, channel uses become independent and the maximum mutual information between input and output during a single channel use defines the channel capacity.
	
\begin{figure}
		\vspace{-0.1cm}
		\begin{center}
			\includegraphics[width=0.75\columnwidth,keepaspectratio]{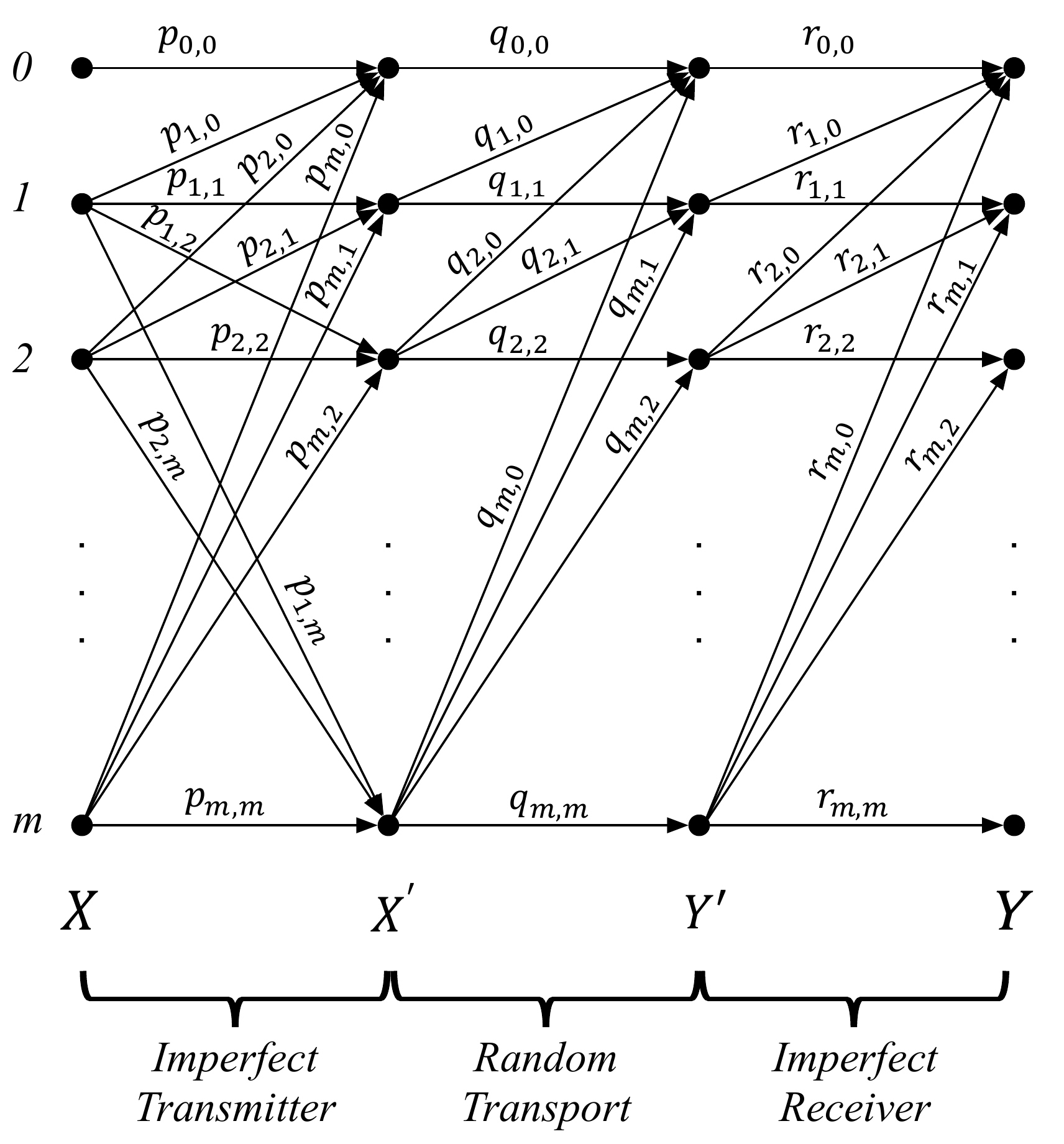}
		\end{center}
		\vspace{-0.35cm}
		\caption{\label{fig:PIchannel} The particle intensity channel with imperfect transmitter.}
		\vspace{-0.5cm}
	\end{figure}

	In practice, particle emission at the transmitter may be stochastic, as we now describe in more detail. Let $X$ be the number of particles that the transmitter {\em intends} to release, and $\Xprim$ be the number of particles that are {\em actually} released into the medium. Moreover, let $p_{x,\xprim}$ be the probability that $\xprim$ particles are actually released when the receiver intended to release $x$ particles (as illustrated in the left portion of Fig.~\ref{fig:PIchannel}). 
	
	We now consider the stochastic particle transport. Each released particle $i$ will arrive at the receiver at some independent identically distributed random time $T_i \sim f_T(\cdot)$.  Let $f_T(t)$ denote the PDF of the time the particle arrives, and $F_T(t)$ denote its corresponding CDF. Then the probability that a particle arrives during a symbol duration is given by	
	\begin{align}
	\label{eq:DefRho}
	\rho = F_T(\tau),
	\end{align}
	and the probability that it never arrives (and is by assumption never detected) is $1-\rho$.
	
	If we assume $F_T(t)$ is strictly monotone, the symbol duration can be obtained from $\rho$ by using the inverse CDF (iCDF) function, i.e., $\tau=F^{-1}_T(\rho)$. Using the iCDF, we can also rewrite $m_\tau$ as a function of $\rho$:
	\begin{align}
	m_\rho  = \left \lfloor \lambda F^{-1}_T (\rho) \right \rfloor
	\end{align}
	
	Let $\Yprim$ denote the total number of particles that arrive at the receiver during the symbol duration. Since we assume 
	the particles propagate independently of each other, given $\xprim$ particles were actually released by the transmitter, we have
	\begin{align}
	\label{eq:PIChan}
	P(\yprim|\xprim;\rho) = \begin{cases}
	q_{\xprim,\yprim} = {\xprim \choose \yprim} \rho^{\yprim} (1-\rho)^{\xprim-\yprim}, & \yprim\leq \xprim\\
	q_{\xprim,\yprim} = 0, & \yprim>\xprim
	\end{cases}
	\end{align}
	and in general $P(\Yprim|\Xprim=\xprim;\rho) \sim \BinomDist(\xprim,\rho)$, where $\BinomDist(n,q)$ indicates the binomial distribution with parameters $n$ and $q$. 
	
	Finally, the receiver may not be able to perfectly detect the particles that arrive. Let $\Yprim$ be the number of particles that arrive at the receiver during one symbol duration, and let $Y$ be the number of particles that are actually detected during the corresponding symbol duration. Furthermore, let $r_{y,\yprim}$ be the probability that $y$ particles are detected by the receiver, when $\yprim$ particles arrive at the receiver. Note that the receiver may fail to detect some of particles that arrive due to sensitivity or uncertainty in the detection process. The end-to-end channel is then defined as 
	%
	%
	%
	%
	\begin{align}
	\label{eq:PIChanTx}
	P(y|x;\rho) = \sum_{j=0}^{m_\rho}\sum_{i=0}^{m_\rho} p_{x,i}q_{i,j}r_{y,j}.
	\end{align}
	Since $\{p_{x,i}\}_{i=0}^{m_\rho}$ is a PMF, one can think of the channel input as actually being this PMF. 
	
	\note{We don't use this matrix notation anywhere, so this last line should be deleted.  HOWEVER, we COULD note that each binomial channel transition matrix $B(a)$ is
		$$
		\left [
		\begin{array}{ccccc}
		1 & 1-a & (1-a)^2 & \cdots & (1-a)^m\\
		0 & a & 2a(1-a) & \cdots & ma (1-a)^{m-1}\\
		\vdots & 0 & a^2 & \cdots & {m\choose 2} a^2 (1-a)^{m-2}\\
		\vdots & 0 & 0 & \ddots & \vdots \\
		0 & \cdots & \cdots & 0 & a^m\\
		\end{array}
		\right ]
		$$
		See all the pretty eigenvalues on the diagonal?  :) Turns out that the eigenvectors are
		$$
		\left [ 
		\begin{array}{c}
		1\\
		0\\
		\vdots\\
		0
		\end{array}
		\right ]
		$$
		$$
		\left [ 
		\begin{array}{c}
		1\\
		-1\\
		0\\
		\vdots\\
		0
		\end{array}
		\right ]
		$$
		$$
		\left [ 
		\begin{array}{c}
		1\\
		-2\\
		1 \\
		0\\
		\vdots\\
		0
		\end{array}
		\right ]
		$$
		$$
		\left [ 
		\begin{array}{c}
		1\\
		-3\\
		3 \\
		-1\\
		0\\
		\vdots\\
		0
		\end{array}
		\right ]
		$$
		etc., none of which have ANYTHING to do with $a$.  Do a $\bf S \Lambda S^{-1}$ factorization and the product theorem pops out directly.
		However, I think the product form proposition and proof are ok.
	}
	\notenf{Good pint. I think we should use this in future extensions of this work.}

	In the rest of this work we assume that 
	\begin{align}
	\label{eq:TxProbs}
	p_{x,\xprim} = {m_\rho \choose \xprim} (\tfrac{\alpha x}{m_\rho})^{\xprim} (1-\tfrac{\alpha x}{m_\rho})^{m_\rho-\xprim},
	\end{align}
	where $0<\alpha\leq1$. Note that here $p_{0,0}=1$, which means the transmitter can send zero particles with perfect accuracy. Moreover, for $\alpha = 1$, the expected number of particles released by the transmitter is $x$, and for $m_\rho$ large, using the Gaussian approximation to a binomial, it is normally distributed around $x$ for $x$ not close to 0 and $m_\rho$. The particles that arrive at the receiver are assumed to be detected independently of each other with probability $\beta$ and therefore we have
	%
	%
	%
	%
	%
	\begin{align}
	\label{eq:RxProbs}
	r_{y,\yprim} = {\yprim \choose y} (\beta)^{y} (1-\beta)^{\yprim-y}.
	\end{align}
	We now present the end-to-end channel characteristic in the following proposition.
	
	\begin{proposition}[Particle-Intensity Channel]
		For an MC system which consists of an imperfect PI transmitter governed by \eqref{eq:TxProbs}, the PI propagation channel in \eqref{eq:PIChan} and an imperfect receiver governed by \eqref{eq:RxProbs}, the channel is characterized by
		\begin{align}
		\label{eq:PIChanTxRx}
		P(Y=y|X=x;\rho) = {m_\rho \choose y} (\tfrac{x \theta_{\rho}}{m_\rho})^{y} (1-\tfrac{x \theta_{\rho}}{m_\rho})^{m_\rho-y},
		\end{align}
		where $\theta_{\rho} \equiv \alpha  \rho \beta$. 
		%
		%
		%
		%
		%
	\end{proposition}
	\begin{IEEEproof}
		Given $X=x$, $\Xprim\sim \BinomDist(m_\rho,\tfrac{x\alpha }{m_\rho})$. We can write this as $\Xprim = \sum_{i=0}^{m_\rho} \Xprim_i$ where $\Xprim_i\sim \BinomDist(1,\tfrac{x\alpha }{m_\rho})$ is an indicator that the \ith particle  is released by the transmitter. Let $Z_i\sim \BinomDist(1,\rho)$ be an indicator that the \ith released particle arrives at the receiver within the symbol duration, and $\Yprim_i = \Xprim_i Z_i$ be the indicator that the \ith particle is both released and arrives at the receiver during the symbol interval. Let $Z^\prime_i\sim \BinomDist(1,\beta)$ be the indicator that the \ith released particle that arrives at the receiver is detected, and $Y_i=\Yprim_i Z^\prime_i $ be the indicator that the \ith particle is released by the transmitter, arrives at the receiver, and is detected. Then $Y_i\sim \BinomDist(1,\tfrac{x\alpha \rho \beta}{m_\rho})$, and since $Y = \sum_{i=0}^{m_\rho}Y_i$, $Y\sim \BinomDist(m_\rho,\tfrac{x\alpha  \rho\beta}{m_\rho})$.
	\end{IEEEproof}
	
	An important observation here is that as the symbol duration $\tau$ changes, $\rho$ changes, and therefore, the channel changes. In this work, we incorporate the optimization of the symbol duration into the formulation of capacity, to present the channel capacity of the memoryless PIC in bits per second. This is one important distinction between this and previous work \cite{sri12,li14,gha15,ami15} where the channel capacity is typically defined in bits per channel use.

	For the case when $m_\rho$ is large and $\theta_{\rho}$ is small, the system can be well approximated by the Poisson distribution \cite{ver69}
	%
	%
	%
	%
	\begin{align}
	\label{eq:PIChanPois}
	P(y|x;\rho) = \frac{(x \theta_\rho)^{y}e^{x \theta_\rho}}{y!}.
	\end{align}
	We write this as $P(y|x;\rho) \sim \PoissDist(x \theta_\rho)$, where $\PoissDist(a)$ indicates the Poisson distribution with parameter $a$. 
	
	\begin{rem}
		Using the Poisson approximation, the PIC in MC systems can be viewed as a more general formulation of the discrete-time Poisson channel used to model optical intensity channels \cite{sha90,lap98,cao14}. The channel input is discrete here and continuous in the discrete-time Poisson channel. Another important difference is that the symbol duration is finite, and the rate of arrival does not scale linearly with the symbol duration. Note that although here we do not consider interfering particles, they can be introduced to the Poisson model in \eqref{eq:PIChanPois} by adding an extra term similar to the dark current in optical communications \cite{sha90,lap98,cao14}. 
		%
		%
		%
		%
	\end{rem}

	\vspace{-0.2cm}
	\subsection{Channel Capacity Formulation}
	\vspace{-0.2cm}
	We now characterize the channel capacity of the PIC. Let $\mathbf{P}_{X,\rho} = [P_{X,\rho}(0),P_{X,\rho}(1),...,P_{X,\rho}(m_\rho)]$ be the channel input PMF and let $\mathcal{P}$ be the set of all valid PMFs. Then the capacity of the channel in \eqref{eq:PIChanTxRx}, in bits per second is defined as
	\begin{align}
	\label{eq:CapPIasP}
	\mathsf{C}(\rho) =  \underset{\mathbf{P}_{X,\rho}\in\mathcal{P}}{\max}  \frac{I(X;Y|\rho)}{F_T^{-1}(\rho)},
	\end{align}
	where $F_T^{-1}(\cdot)$	is the iCDF of  the particle detection time. 
	Since the channel changes as a function of the symbol duration, the fundamental limit of this channel is then
	\begin{align}
	\label{eq:capPI}
	\mathsf{C}^*=\underset{\rho}{\max}~\mathsf{C}(\rho).
	\end{align}
	
	\begin{rem}
		Note we define the capacity in terms of bits per second, as the capacity or the rate of a particle's arrival typically does not increase linearly with time. This is in contrast to the discrete-time Poisson channel in optical communications, where the rate of photon arrival increases linearly with the symbol duration \cite{cao14}. Therefore, in the PIC, symbol duration can have a significant effect on the information rate. This will be further demonstrated in Section \ref{sec:results}.
		%
		%
		%
		%
	\end{rem}
	
	\vspace{-0.2cm}
	\section{Optimal Input Distribution \\ and Binary-Input Capacity}
	\label{sec:OptimalInputBCap}
	\vspace{-0.2cm}
	We now investigate the characteristics of the optimal input distribution, present capacity when the input is binary, and investigate  settings for which a capacity-achieving distribution is binary.
	
	Let $\mathbf{P}_{X,\rho}^*$ be the optimal input PMF that maximizes the mutual information in \eqref{eq:CapPIasP}, for a given $\rho$, and let $P_{X,\rho}^*(i)$ be the \ith element of $\mathbf{P}_{X,\rho}^*$. The following theorem shows that the input distribution $\mathbf{P}_{X,\rho}^*$ that maximizes the mutual information in \eqref{eq:CapPIasP} has mass points at 0 and $m_\rho$.
	
	\begin{theorem}
		\label{thm:massPoints}
		For a given symbol duration $\tau$, and hence a given $\rho$, the mutual information given in \eqref{eq:CapPIasP} is maximized by a PMF $\mathbf{P}_{X,\rho}^*$, where $P_{X,\rho}^*(0)>0$ and $P_{X,\rho}^*(m_\rho)>0$. 
	\end{theorem}
	
	\begin{IEEEproof}
		First, we prove that $P_{X,\rho}^*(0)>0$ by contradiction. The mutual information in \eqref{eq:CapPIasP} can be written as 
		\begin{align*}
		I(X;Y|\rho) = H(X|\rho) - H(X|Y,\rho).
		\end{align*} 
		Assume that $P_{X,\rho}^*(0)=0$ in $\mathbf{P}_{X,\rho}^*$, and let $X^*$ be a RV drawn from this PMF, and $Y^*$ the corresponding channel output. 
		Let $0<i<m_\rho$ be the \ith~index  of $\mathbf{P}_{X,\rho}^*$ such that $P_{X,\rho}^*(i)>0$. 
		Let $\mathbf{P}_{X,\rho}^\dagger$ be another input distribution constructed by swapping the \ith element in $\mathbf{P}_{X,\rho}^*$ with the zeroth element. That is $P_{X,\rho}^\dagger(0)=P_{X,\rho}^*(i)$, $P_{X,\rho}^\dagger(i)=0$, and all the other elements of $\mathbf{P}_{X,\rho}^\dagger$ are the same as $\mathbf{P}_{X,\rho}^*$. Let $X^\dagger$ be a RV drawn from this PMF, and $Y^\dagger$ the corresponding channel output. Clearly $H(X^*|\rho)=H(X^\dagger|\rho)$. Since symbol 0 is always transmitted perfectly, as there is no ISI and interfering particles, $H(X^\dagger|Y^\dagger,\rho)\leq H(X^*|Y^*,\rho)$. Therefore we have $I(X^*;Y^*|\rho)\leq I(X^\dagger;Y^\dagger|\rho)$,
		which contradicts our assumption that $\mathbf{P}_{X,\rho}^*$ is optimal. Therefore, the optimal input distribution always satisfies $P_{X,\rho}^*(0)>0$. Using the same reasoning, and the fact that between all the non-zero symbols, the symbol $m_\rho$ results in the highest probability that a particle arrives at the receiver, we can prove that the optimal input distribution always satisfies $P_{X,\rho}^*(m_\rho)>0$ as well. 
	\end{IEEEproof}	
	
	Previous work on MC has considered on-off-keying in diffusive environments \cite{far16ST}. We now derive the capacity in \eqref{eq:CapPIasP} as a function of $\rho$ for this particular class of MC systems. 
	
	We assume that the particles that arrive at the receiver are either immediately detected by the receiver with probability $\beta$, or they are never detected, perhaps denatured as part of misdetection. Let $\ell$ be the shortest distance between a point source transmitter and the surface of a spherical receiver, $d$ be the diffusion coefficient of particles, and $r$ be the radius of a spherical receiver. If we assume that the particles diffuse from the transmitter, located at the origin, to the spherical receiver, the motion of each particle can be represented using a random Brownian path in 3D space. Since we assume that the particles are either detected when they arrive at the receiver or they are never detectable, the time of arrival is given by the first time the particle reaches the receiver. For Brownian motion in 3D space, the first arrival time $T$, to the spherical receiver, is a scaled L\'evy-distributed random variable where the scale constant is $\eta=\tfrac{r}{\ell+r}$ \cite{yilmaz20143dChannelCF}. This means that there is a non-zero probability that a particle never arrives at the receiver. Note that for Brownian motion in 1D space, $\eta=1$.  Using the iCDF of the scaled \levy distribution, we obtain
	\begin{align}
	\label{eq:tauDiff}
	\tau = F_T^{-1}(\rho) =\frac{c}{2\erfcinv^2(\rho/\eta)}.		
	\end{align}
	where $c=\tfrac{\ell^2}{2d}$ and $\erfcinv(.)$ is the inverse of the complementary error function $\erfc(.)$. We call a channel that relies on this diffusive transport the {\em diffusion-based PIC (DBPIC)}. 
	%
	%
	%
	%
	\begin{rem}
		Substituting \eqref{eq:tauDiff} into \eqref{eq:CapPIasP}, we observe that the diffusion coefficient $d$ has no effect on the optimal input distribution and the optimal $\rho$. This is despite the fact that the capacity increases linearly with $d$. This means that if the type of particle is changed, so long as the distance between the transmitter and the receiver is the same, and the receiver has the same radius, the optimal distribution and the optimal $\rho$ values will remain the same. Note that the change in capacity is due to the fact that a shorter or a longer symbol duration is required to achieve the same $\rho$ value.
		%
		%
		%
		%
	\end{rem}
	
	\begin{rem}
		If we consider a 1D environment\footnote{Note that a 1D environment is a good approximation if the system is confined inside a very narrow and long physical channel.} (i.e., $\eta=1$), we observe that the capacity decreases as $\tfrac{1}{l^2}$, and the distance $l$ does not affect the optimal input distribution and the optimal $\rho$. For a 3D environment however, changing the distance $l$ and the radius $r$ could affect the optimal $\rho$ and $P(x)$ values through $\eta$. 
		%
		%
		%
		%
	\end{rem}

	We now present the channel capacity of  the binary-input DBPIC in the following theorem.
	%
	%
	%
	%
	\begin{theorem}
		\label{thm:binaryCap}
		Let $X^b \in \{0,m_\rho\}$ be the binary input to the DBPIC in \eqref{eq:PIChanTxRx}, $\xi_\rho$ be the probability that the symbol $m_\rho$ is transmitted, and $\varphi_\rho=(1-\theta_\rho)^{m_\rho}$. Then an optimal input distribution $\xi^*_\rho$ is given by 
		%
		%
		%
		%
		%
		%
		%
		%
		\begin{align}
		\label{eq:binOptiInput}
		\xi^*_\rho=\frac{1}{\varphi_\rho^{\tfrac{\varphi_\rho}{\varphi_\rho-1}}-\varphi_\rho+1}, 
		\end{align} 
		and the capacity of \eqref{eq:CapPIasP}, in bits per second, is given by
		\begin{align}
		\label{eq:binCapCEMperP}
		\mathsf{C}_{DB}^b(\rho)=\tfrac{2}{c}\erfcinv^2(\rho/\eta)\log\left( 1+(1-\varphi_\rho)\varphi_\rho^{\tfrac{\varphi_\rho}{1-\varphi_\rho}}\right).
		\end{align}	  
	\end{theorem}
	
	\begin{IEEEproof}
		For the binary input DBPIC, the mutual information in \eqref{eq:CapPIasP} can be written as a function of $\xi_\rho$ using
		\begin{align}
		\label{eq:mutualBinary}
		I(X^b;Y|\rho) = I(\xi_\rho) = &\xi_\rho\varphi_\rho\log(\varphi_\rho)-\xi_\rho(1-\varphi_\rho)\log(\xi_\rho) \nonumber\\
		&-(1-\xi_\rho+\xi_\rho\varphi_\rho)\log(1-\xi_\rho+\xi_\rho\varphi_\rho).
		\end{align}
		Taking the derivative of $I(\xi_\rho)$ with respect to $\xi_\rho$, setting it equal to zero, and solving the resulting equation, we obtain \eqref{eq:binOptiInput}. Substituting \eqref{eq:binOptiInput} into \eqref{eq:mutualBinary} and using \eqref{eq:tauDiff} in \eqref{eq:CapPIasP} we obtain the capacity expression in \eqref{eq:binCapCEMperP}. 
	\end{IEEEproof}	
	The log expression in the capacity result in Theorem \ref{thm:binaryCap} is similar to the capacity of the z-channel because, for a binary input, the channel reduces to the z-channel with symbol 0 going to 0 with probability 1, and symbol $m_\rho$ going to symbol 0 with probability $\varphi_\rho$ \cite{tal02}.
	%
	%
	%
	%

	An interesting question that arises here is the following: when is the binary input optimal for the PIC. In the following proposition, we provide a guideline for the optimality of the binary input for a  subclass of PICs.
	\begin{proposition}
		\label{prop:CondiBinary}
		For the PIC in \eqref{eq:PIChanTxRx} where $m_\rho$ is large and $\theta_{\rho}$ is small such that the Poisson approximation in \eqref{eq:PIChanPois} is accurate, the binary input distribution given in \eqref{eq:binOptiInput} is optimal if $m_\rho \theta_{\rho} < 3.3679$.
	\end{proposition}
	\begin{IEEEproof}
		Using the same technique presented in \cite{sha90} for the optical channels, the proposition can be proved. 
	\end{IEEEproof}	
	
	Note that this condition may be satisfied in many practical systems where the radius of the receiver is much smaller than the distance between the transmitter and the receiver, hence the probability of particles arriving is small, and the rate of particle generation is small. Upper bounds on the total variation between binomial and Poisson distributions can be used to show that this variation is small for small $\theta_{\rho}$ \cite{ver69}.      
	%
	%
	%
	%
	%
	
	\begin{figure}
		\begin{center}
			\includegraphics[width=0.95\columnwidth,keepaspectratio]{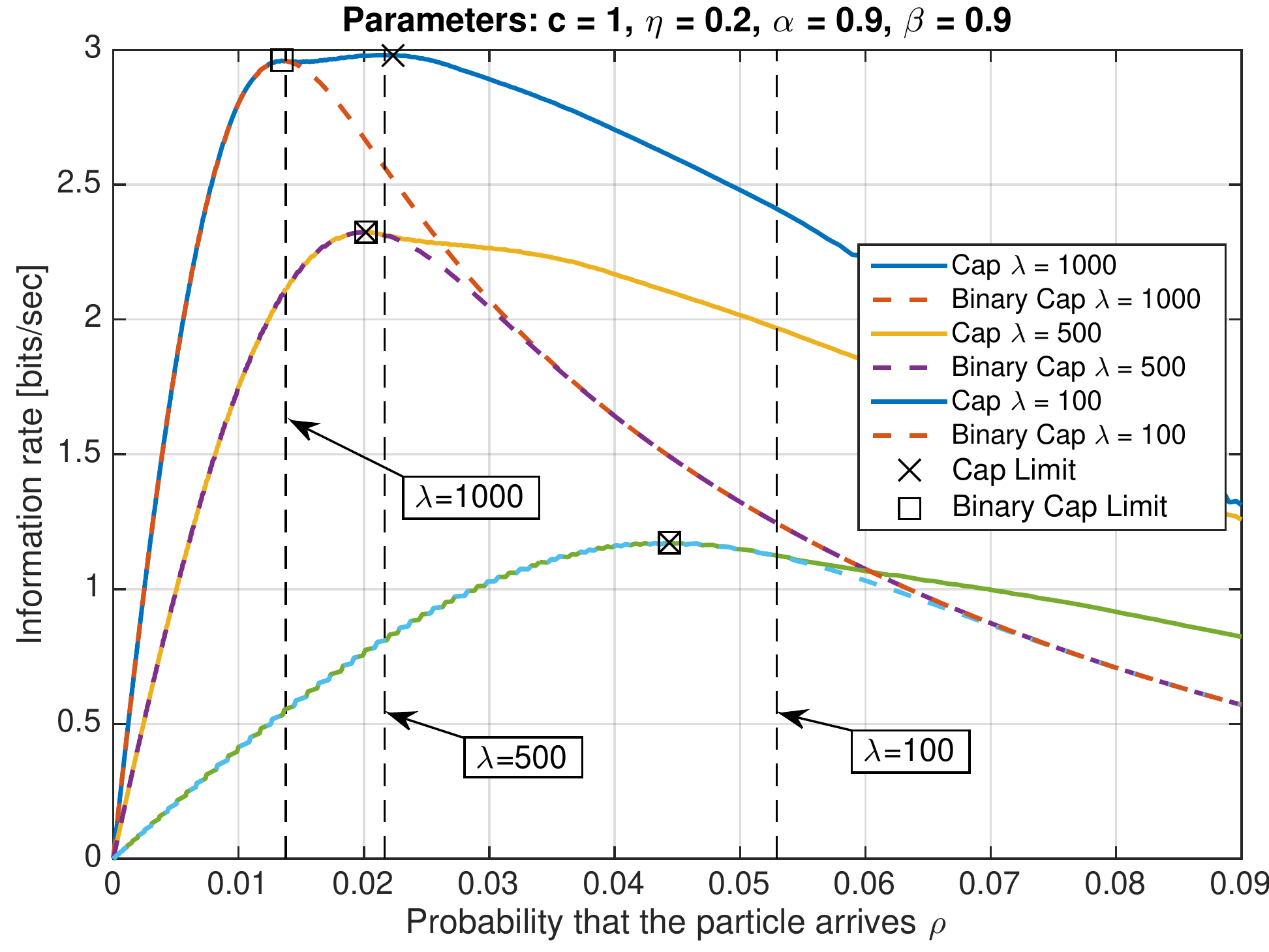}
		\end{center}
		\vspace{-0.4cm}
		\caption{\label{fig:capPlot} The information rate for binary-input (dashed lines) and the optimal input based on Blahut-Arimoto algorithm (solid lines) for different $\lambda$. The three vertical dashed lines indicate the $\rho$ value after which $m_\rho \theta_{\rho}$ in Proposition \ref{prop:CondiBinary} is greater than 3.3676.}
		\vspace{-0.4cm}
	\end{figure}

	\section{Numerical Results}
	\label{sec:results}
	
\begin{figure}
		\begin{center}
			\includegraphics[width=0.95\columnwidth,keepaspectratio]{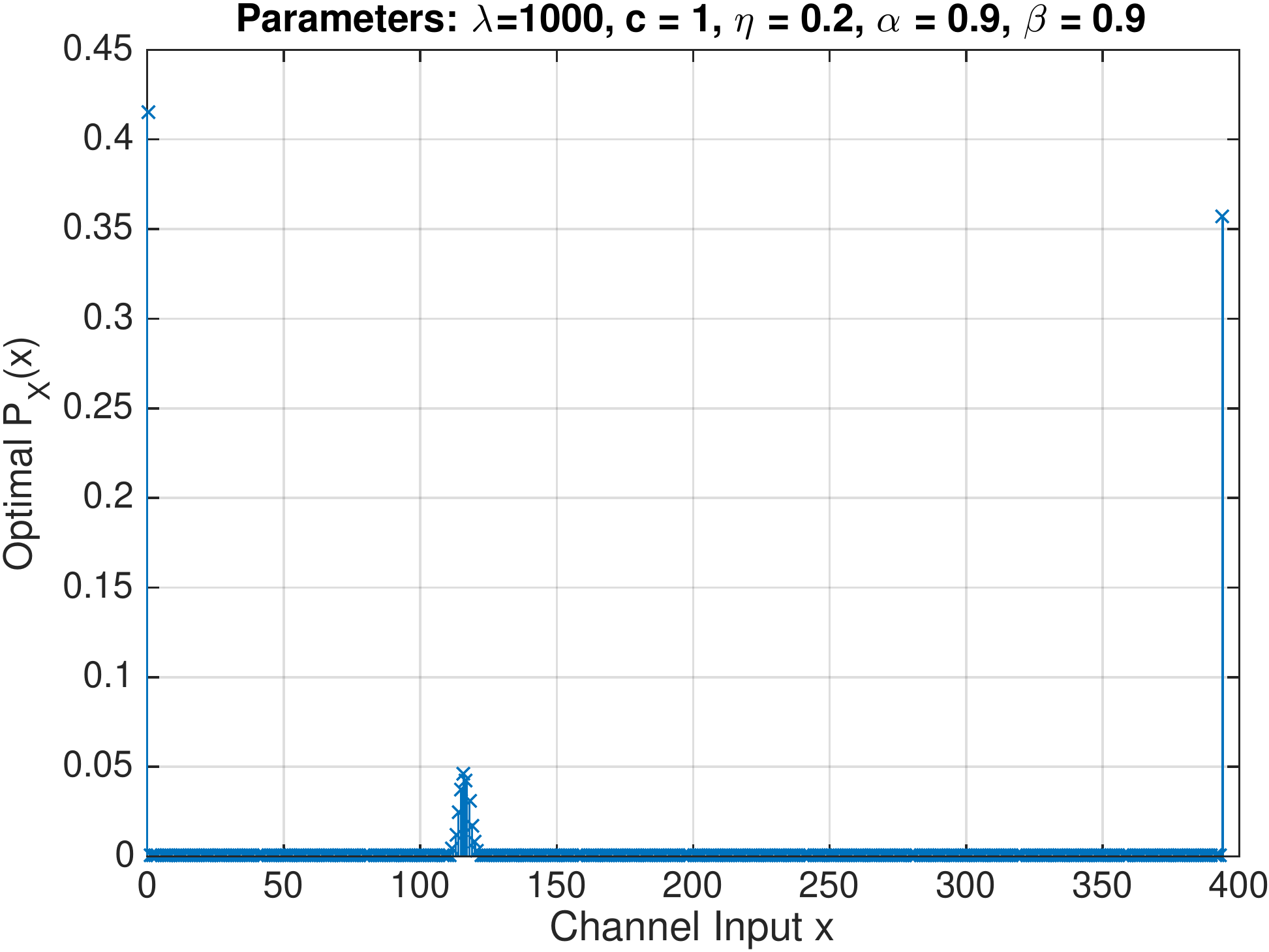}
		\end{center}
		\vspace{-0.4cm}
		\caption{\label{fig:OptimalInputPlot} The optimal input distribution that maximizes $\mathsf{C}^*$ for $\lambda=1000$.}
		\vspace{-0.5cm}
	\end{figure}

	We now numerically compare capacity for an optimal input distribution against capacity under the binary distribution for DBPIC. For the comparison, the optimal input distribution and the capacity are calculated using the Blahut-Arimoto algorithm~\cite{bla72}. Fig.~\ref{fig:capPlot} shows the results for three different particle generation rates, $\lambda$. The scaling factor for the \levy distribution $\eta=0.2$, which means the distance between the transmitter and the receiver is four times the radius of the receiver. The square markers indicated the $\mathsf{C}^*$ in \eqref{eq:capPI} for the binary input distribution, and the X-markers indicated $\mathsf{C}^*$ for the optimal input distribution. Based on the results we observe that it is only for the case of $\lambda=1000$ that the binary input distribution does not maximize $\mathsf{C}^*$. The three vertical dashed lines indicate the $\rho$ value after which $m_\rho \theta_{\rho}$ in Proposition \ref{prop:CondiBinary} is greater than 3.3676. We observe that for the $\rho$ values smaller than this critical value, the binary input is the optimal input distribution. Fig.~\ref{fig:OptimalInputPlot} shows the optimal input distribution that maximizes $\mathsf{C}^*$ in \eqref{eq:capPI} for $\lambda=1000$.    
	%
	%
	%
	%
	
	\vspace{-0.15cm}
	\section{Conclusions}
	\label{sec:conclusion}
	\vspace{-0.15cm}
	We introduce the PIC and show that an optimal input distribution for this channel always has mass points at zero and the maximum number of particles that can be released by the transmitter. Interestingly, it was observed that for diffusion-based propagation, the diffusion coefficient, and hence the type of the particles used, does not affect the optimal input distribution. We then derive capacity for the binary input diffusion-based PIC and present conditions under which a binary input is optimal for this channel. Our numerical results illustrate that a binary input is optimal for systems where the transmitter cannot generate particles at rates that satisfy Proposition \ref{prop:CondiBinary}. This can be thought of as the low SNR regime. As part of future work, we will introduce ISI into our model.
	%
	%
	%
	%
	%
	%

	\tiny
	\bibliographystyle{IEEEtran}
	\bibliography{IEEEabrv,MolCom}
	

\end{document}